\documentclass[pdftex,singlecolumn,epjc3]{svjour3}
\RequirePackage{graphicx}
\RequirePackage{mathptmx}      
\RequirePackage{flushend}
\RequirePackage[numbers,sort&compress]{natbib}
\RequirePackage[colorlinks,citecolor=blue,urlcolor=blue,linkcolor=blue]{hyperref}
\usepackage{amsmath}
\usepackage{graphicx}
\usepackage{geometry}
\usepackage{caption}
\usepackage{subcaption}
\usepackage{color}
\usepackage{orcidlink}
\begin{document}
\UseRawInputEncoding

\journalname{IJMPA}

\title{Rotational and inverse square potential effects on harmonic oscillator confined by flux field in a space-time with screw dislocation }

\author{Faizuddin Ahmed\orcidlink{0000-0003-2196-9622}\,\thanksref{addr1,e1} \and Houcine Aounallah\orcidlink{0000-0002-0611-4063}\thanksref{addr2,e2} \and Prabir Rudra\orcidlink{0000-0002-5791-5696}\thanksref{addr3,e3}
}

\thankstext{e1}{e-mail: \bf faizuddinahmed15@gmail.com; faizuddin@ustm.ac.in}
\thankstext{e2}{e-mail: \bf houcine.aounallah@univ-tebessa.dz}
\thankstext{e3}{e-mail: \bf prudra.math@gmail.com (corresponding author)}

\institute{{Department of Physics, University of Science \& Technology Meghalaya, Ri-Bhoi, Meghalaya-793101, India}\label{addr1}
\and {Department of Science and Technology, Echahid Cheikh Larbi Tebessi University - Tebessa, Algeria}\label{addr2} \and {Department of Mathematics, Asutosh College, Kolkata-700026, India}\label{addr3}
}

\vspace{0.75cm}

\date{\today}


\vspace{0.75cm}

\maketitle

\abstract{This research paper delves into the study of a non-relativistic quantum system, considering the interplay of non-inertial effects induced by a rotating frame and confinement by the Aharonov-Bohm (AB) flux field with potential in the backdrop of topological defects, specifically a screw dislocation. We first focus on the harmonic oscillator problem, incorporating an inverse-square repulsive potential. Notably, it becomes evident that the energy eigenvalues and wave functions are intricately influenced by multiple factors: the topological defect parameter $\beta$ (representing the screw dislocation), the presence of a rotating frame engaged in constant angular motion with speed $\Omega$, and the external potential. Then we study the quantum behavior of non-relativistic particles, engaging in interactions governed by an inverse square potential, all while taking into account the effects of the rotating frame. In both scenarios, a significant observation is made: the quantum flux field's existence brings about a shift in the energy spectrum. This phenomenon bears a resemblance to the electromagnetic Aharonov-Bohm effect.
}

\vspace{0.5cm}


\vspace{0.5cm}


\section{Introduction}

It is well known that in the primordial universe, during the phase transition process, there was a decoupling of the fundamental interactions which was associated with symmetry breaking. Due to this, topological defects \cite{defects} are expected to be observed in the universe in the form of exotic cosmological objects. Some of these are domain wall \cite{domain}, cosmic string \cite{string1, string2, string3}, global monopole (GM) \cite{monopole}, etc. Since these objects are yet to be observed, they can still be considered hypothetical objects, but some of them are so obvious that they are expected to be observed in near future. GM is the most promising one that is expected to be observed, and hence, a lot of research has been conducted on it in recent years \cite{gm2, gm3, gm4, gm5, gm6, gm7}. GM has also been investigated in the arena of relativistic quantum mechanics on the hydrogen atom and on the pionic atom \cite{qm1}. The exact solutions of the Klein-Gordon equation have been extensively investigated for GM, especially in the presence of dyon, magnetic flux, and scalar potential \cite{gm8}. Moreover, thorough investigations on GM have been conducted in the exact solutions of scalar bosons with the Aharonov-Bohm effect and Coulomb potential \cite{gm9, gm10, gm11}. In addition to this, in the context of Dirac and Klein-Gordon oscillators, GM has been extensively studied \cite{gm12, gm14, gm15}. Further work on the topological defects can be found in \cite{def2, def3, def4}. In addition, cosmic string in the context of the relativistic wave equations without or with external magnetic field have been studied by many authors in the literature.

In quantum mechanics, various studies have been undertaken which involve non-inertial effects, and these probes have uncovered interesting features in interferometry associated with geometric phases \cite{ni1, ni2, ni3, ni4}. These studies have also brought into light a new type of coupling between the angular momentum and the angular velocity of the rotating frame \cite{ni5, ni6, ni7}. There are other studies in literature dealing with geometric phases in non-inertial systems made via Lorentz transformations \cite{ni8}, weak field approximation \cite{ni9}, and via the analogue effect of the Aharonov-Casher effect \cite{ni10}. Non-inertial effects have also been explored in confined quantum systems such as Landau-Aharonov-Casher quantization \cite{ni11}, and Dirac oscillator \cite{ni12}. Investigations have also been carried out in the confinement of a neutral particle to a quantum dot \cite{ni13, ni14}. Of late the study of non-inertial effects has been expanded to the quantum dynamics of the neutral particles possessing a permanent magnetic field and electric dipole moments. In the Fermi-Walker reference frame, a neutral particle with a permanent magnetic dipole moment has been explored \cite{ni15}. In a curved space-time, non-inertial effects of the Fermi-Walker reference frame have been investigated in the presence of holonomies \cite{ni16}.

Furthermore, the non-relativistic wave equation in the background of topological defects with potential of various kinds has been investigated in the literature. A few well-known potential models are pseudoharmonic potential, Mie-type potential, Kratzer potential, Yukawa potential, Hulthen potential, Morse potential, hyperbolic Potential, non-central potential, Manning-Rosen (MR) potential, Rosen-Morse potential, Deng-Fan potential, Poschl-Teller potential, etc. (see, Refs. \cite{gg1,gg2,gg3,gg4,gg5,gg6,gg7,gg8,gg9,gg10,FA2,FA4,FA5,FA7} and related references therein). To obtain the eigenvalue solutions of the non-relativistic wave equation, various methods or techniques have been employed by many authors in the literature, such as the parametric Nikiforov-Uvarov method and its functional analysis, Supersymmetric quantum mechanics (SUSY), Asymptotic iteration method (AIM), Laplace's transformation method, through confluent hypergeometric and biconfluent Heun equations method, etc. and many more. Noted that all these investigations have been carried out in flat space background in four- and higher dimensions.

In this particular analysis, we study the harmonic oscillator problem in the presence of an inverse square potential given by \cite{FC,RS,RR,SMI,AC,SMI2}
\begin{equation}
V (r)=\frac{\gamma}{r^2}+\delta,
\label{aa1}
\end{equation}
where $\delta, \gamma$ are constants.

Our analysis is motivated by the findings outlined in Ref. \cite{LCNS}, where the authors examined the impact of non-inertial effects on a non-relativistic quantum system governed by a harmonic oscillator in a space-time environment featuring a screw dislocation. Building upon these insights, we extend our investigation to explore the quantum system of a harmonic oscillator under the influence of non-inertial effects induced by a rotating frame. This exploration includes scenarios where the harmonic oscillator experiences a centrifugal potential or an inverse square potential, all within the context of a quantum flux field and the same topological defect geometry characterized by a screw dislocation.

The core of our study involves solving the radial wave equation using the confluent Heun equation, enabling us to attain analytical solutions for the eigenvalues of the harmonic oscillator. Notably, our analysis reveals that the energy levels and wave functions of the harmonic oscillator are notably influenced by the combined effects of the centrifugal potential and the presence of the quantum flux field. These factors collectively lead to modifications in the energy spectrum and wave functions compared to the outcomes documented in Ref. \cite{LCNS}. Subsequently, our investigation extends to the realm of non-relativistic particles engaged in interactions with either a centrifugal potential or an inverse square potential, all while subject to the influence of the quantum flux field within the same topological defect geometry. By solving the radial wave equation and employing our established procedure, we secure the eigenvalue solutions for this scenario.

This paper is structured as follows: In Section 2, we delve into the study of a harmonic oscillator confined by a quantum flux field within a space-time characterized by a topological defect featuring a screw dislocation. The confinement is further compounded by the presence of an inverse square potential. Our analysis leads us to derive the eigenvalue solution for this intricate quantum system. In Section 3, our focus shifts to the quantum system of non-relativistic particles. Here, we consider the interaction of particles exclusively with an inverse square potential, all while being subject to the influence of the quantum flux field. We rigorously solve the underlying radial wave equation and extract the corresponding eigenvalue solution. Finally, in Section 4, we present our overarching conclusions. These findings encapsulate the essence of our study, highlighting the novel insights gained throughout our analysis.

\section{Rotational effects on harmonic oscillator with potential under flux field in topological defect geometry}

In this section, we investigate the quantum motions of harmonic oscillator problem confined by the quantum flux field under rotational frame effects in the presence of an inverse square potential in a space-time background with a screw dislocation and analyze the eigenvalue solution.

Thereby, we start this section with the Hamiltonian operator of quantum system with interaction potential $V(r)$ described by \cite{LCNS,FA6} ($c=1=\hbar=G$)
\begin{eqnarray}
\hat{H}_0=-\frac{1}{2\,M}\,\Bigg[\frac{1}{\sqrt{g}}\,D_{i}\,\Big(\sqrt{g}\,g^{ij}\,D_{j}\Big)\Bigg]+\frac{1}{2}\,M\,\omega^2_{0}\,r^2+V(r),
\label{aa2}
\end{eqnarray}
where $M$ is the mass of the non-relativistic particles, $D_{i} \equiv (\partial_{i}-i\,e\,A_{i})$ with $e$ is the electric charges, and $A_i$ is the electromagnetic three-vector potential. In this analysis, we choose the electromagnetic three-vector potential $\vec{A}$ with only the non-zero component given by \cite{RLLV}
\begin{equation}
A_r=0=A_z,\quad A_{\phi}=\frac{\Phi_{AB}}{2\,\pi}=\frac{\Phi}{e},
\label{aa}
\end{equation}
where $\Phi_{AB}=const$ is the Aharonov-Bohm magnetic flux, $\Phi_0=2\,\pi\,e^{-1}$ is the quantum of magnetic flux, and $\Phi=\frac{\Phi_{AB}}{\Phi_0}$ is the amount of magnetic flux which is a positive real number. It is worth mentioning that the Aharonov–Bohm effect has investigated in different branches of physics by many authors (see, Ref. \cite{gm15} and related references therein).

The topological defect geometry under consideration in the coordinates $(r, \phi, z)$ is described by \cite{KV,LCNS}
\begin{equation}
ds^2_{3D}=dr^2+r^2\,d\phi^2+2\,\beta\,d\phi\,dz+dz^2,
\label{aa3}
\end{equation}
where the parameter $\beta$ is a positive constant that characterises the screw dislocation (torsion field) \cite{ee1,ee2,ee3}. The ranges of this dislocation parameter are in the intervals $0 < \beta <1$. The metric tensor $g_{ij}$ and its contravariant form $g^{ij}$ are given by
\begin{eqnarray}
g_{ij}=\left(\begin{array}{ccc}
1 &0  &0  \\
0 &r^2 &\beta\\
0 &\beta  &1
\end{array}\right),\quad g^{ij}=\frac{1}{(r^2-\beta^2)}\left(\begin{array}{ccc}
r^2-\beta^2 &0  &0  \\
0 &1 & -\beta\\
0 &-\beta  & r^2
\end{array}\right).
\label{matrix}
\end{eqnarray}
with the determinant of the metric tensor given by $g=|g_{ij}|=(r^2-\beta^2)$. It is worth mentioning that the non-relativistic particles in the presence of screw dislocation has been studied in Refs. \cite{ff1,ff2,ff3}.

Moreover, we want to analyze the rotating frame effects in the quantum system. Let $\Omega$ is a constant angular velocity of the rotating frame, $\vec{\Omega}=\Omega\,\hat{z}$. Therefore, the Hamiltonian operator of the system under a uniform rotation will be \cite{LCNS}
\begin{equation}
H=H_0-\vec{\Omega}\bullet \vec{L},
\label{aa4}
\end{equation}
where $\vec{L}$ is the angular momentum operator.

One can show easily that the presence of a screw dislocation will change the $z$-component of the angular momentum operator given by
\begin{equation}
\vec{L}^{eff}_z=-i\,\Bigg(\frac{\partial}{\partial\,\phi}-\frac{i\,e\,\Phi_{AB}}{2\,\pi}-\beta\,\frac{\partial}{\partial\,z}\Bigg)\,\hat{z}.
\label{aa5}
\end{equation}

Writing Eq. (\ref{aa4}) using (\ref{aa2}) in the space-time background (\ref{aa3}), we have
\begin{eqnarray}
&&-\frac{1}{2\,M}\,\Bigg[\frac{\partial^2}{\partial\,r^2}+\Big(\frac{r}{r^2-\beta^2}\Big)\,\frac{\partial}{\partial\,r}+\frac{1}{(r^2-\beta^2)}\,\Big(\frac{\partial}{\partial\,\phi}-i\,\Phi-\beta\,\frac{\partial}{\partial\,z}\Big)^2 \Bigg]\,\Psi+i\,\Omega\,\Big(\frac{\partial}{\partial\,\phi}-i\,\Phi-\beta\,\frac{\partial}{\partial\,z}\Big)\,\Psi\nonumber\\
&&+\frac{1}{2}\,M\,\omega^2_{0}\,r^2\,\Psi+V(r)\,\Psi=E\,\Psi.
\label{aa6}
\end{eqnarray}

By the method of separation of variables, one can express the wave function $\Psi (r, \phi, z)=\psi(r)\,f(\phi)\,h(z)$. Let us choose a possible solution given by
\begin{equation}
\Psi(r, \phi, z)=e^{i\,\ell\,\phi}\,e^{i\,k\,z}\,\psi (r),
\label{aa7}
\end{equation}
where $\ell=0,\pm\,1,\pm\,2,\pm\,3,.....$ are the eigenvalues of the angular momentum operator and $k > 0$ is a constant.

Therefore, substituting potential (\ref{aa1}) and the wave function Eq. (\ref{aa7}) into the Eq. (\ref{aa6}), we obtain
\begin{equation}
\psi''+\Big(\frac{r}{r^2-\beta^2}\Big)\,\psi'+\Bigg[\Lambda-M^2\,\omega^2_{0}\,r^2-\frac{2\,M\,\gamma}{r^2}-\frac{\iota^2}{(r^2-\beta^2)}\Bigg]\psi=0,
\label{aa8}
\end{equation}
where we have set the parameters
\begin{eqnarray}
\Lambda=2\,M\,(E+\Omega\,\iota-\delta)-k^2,\quad \iota=(\ell-\Phi-\beta\,k).
\label{aa9}
\end{eqnarray}

Performing a change of variable $x=\frac{r^2}{\beta^2}$ in Eq. (\ref{aa8}), we obtain the following differential equation:
\begin{eqnarray}
&&4\,x\,\psi''(x)+\frac{(4\,x-2)}{(x-1)}\,\psi'(x)+\Bigg[\Lambda\,\beta^2-\omega^2\,x-\frac{2\,M\,\gamma}{x}-\frac{\iota^2}{x-1}\Bigg]\,\psi(x)=0,
\label{aa10}
\end{eqnarray}
where we set the parameter $\omega=M\,\omega_0\,\beta$, and $\omega_0$ is the oscillator frequency.

The requirement of the wave-function tells us that the wave-function $\psi(x)$ must be regular everywhere for $x \to 0$ and $x \to \pm\,\infty$. Let us suppose a possible solution to the Eq (\ref{aa10}) given by
\begin{equation}
\psi(x)=x^{\frac{1}{4}+\frac{j}{2}}\,e^{-\frac{1}{2}\,\omega\,x}\,G(x),
\label{aa11}
\end{equation}
where $j=\sqrt{2M\gamma+\frac{1}{4}}$ and $G(x)$ is an unknown function of $x$.

Substituting this solution (\ref{aa11}) into the Eq. (\ref{aa10}), one will arrive at the second-order homogeneous confluent Heun differential equation form \cite{AR,AR2,AR3}. Thus, the function $G(x)$ is none other than the confluent Heun function which is in our case given by
\begin{equation}
G(x)=H_{c}\left(-\omega,j,-\frac{1}{2},\frac{\Lambda\beta^{2}}{4},\frac{3}{8}-\frac{\iota^{2}+\Lambda\beta^{2}}{4};x\right).
\label{aa12}
\end{equation}

To solve the differential equation (\ref{aa10}), we assume the confluent Heun function $G(x)$ to be a power series solution around the origin given by \cite{GBA}
\begin{equation}
G(x)=\sum_{i=0}^{\infty}\,c_{i}\,x^{i}.
\label{aa13}
\end{equation}
Thereby, substituting this power series (\ref{aa13}) in the Eq. (\ref{aa12}), we obtain the following recurrence relation
\begin{equation}
c_{i+2}=\frac{1}{d_3}\,\Big(d_1\,c_{i+1}+d_2\,c_{i}\Big)
\label{aa14}
\end{equation}
with the coefficient
\begin{equation}
c_{1}=\frac{\Big[2\,\omega\,(1+j)-\iota^{2}-\Lambda\beta^{2}+\frac{1}{2}+j\Big]}{4\,(1+j)}\,c_{0},
\label{aa15}
\end{equation}
where we have set the parameters
\begin{eqnarray}
&&d_{1}=\left(i+\omega+\frac{3}{2}+j\right)\left(i+1\right)-\frac{\iota^{2}+\Lambda\beta^{2}-\frac{1}{2}-j-2\,\omega\,(1+j)}{4},\nonumber\\
&&d_{2}=-\omega\,i+\frac{\Lambda\beta^{2}-\omega\,(3+2\,j)}{4},\quad d_{3}=\left(i+\frac{3+2\,j}{2}\right)\left(i+2\right).
\label{aa18}
\end{eqnarray}

One can see from the power series solution (\ref{aa13}) that it must be a finite degree polynomial of degree $n$ so that the wave-function $\psi(x)$ considered in the Eq. (\ref{aa11}) is finite everywhere. For that, let us consider the case $i=(n-1)$ when the coefficient $c_{n+1}=0$, we obtain from the recurrence relation (\ref{aa15}) the following relation
\begin{eqnarray}
c_{n}=\frac{\Big[4\,\omega\left(n-1\right)-\Lambda\beta^{2}+2\,\omega\,\Big(j+\frac{3}{2}\Big)\Big]}{\Big[4\,n\,(n+\omega+\frac{1}{2}+j)-\iota^{2}-\Lambda\,\beta^{2}+\frac{1}{2}+j+2\,\omega\,(1+j)\Big]}\,c_{n-1}.
\label{aa19}
\end{eqnarray}

Now, it is convenient to evaluate the individual energy levels and the wave functions of harmonic oscillator problem. As particular case, we consider the lowest state of the quantum system defined by the radial mode $n=1$, we obtain from (\ref{aa19}) the coefficient
\begin{equation}
c_{1}=\Bigg[\frac{\omega\,(3+2\,j)-\Lambda\,\beta^{2}}{6+2\omega\,(j+3)+5\,j-\iota^{2}-\Lambda\beta^{2}+\frac{1}{2}}\Bigg]\,c_{0}.
\label{aa20}
\end{equation}

Comparing Eq. (\ref{aa15}) with Eq. (\ref{aa20}) and finally using Eq. (\ref{aa9}), we obtain the ground state energy levels $E_{1,\ell}$ satisfying the following relation
\begin{equation}
E_{1,\ell}=\frac{k^{2}}{2\,M}+\frac{\lambda}{2\,M}+\delta-\Omega\,(\ell-\Phi-\beta\,k),
\label{aa21}
\end{equation}
where
\begin{eqnarray}
&&\lambda=\frac{1}{\beta^2}\,\Bigg[3-2\,(\ell-\Phi-\beta\,k)^2+4\,M\,\omega_0\,\beta\,(2+j)+2\,j \pm \sqrt{\triangle}\Bigg],\quad j=\sqrt{2\,M\,\gamma+\frac{1}{4}},\nonumber\\
&&\triangle=16\,(\ell-\Phi-\beta\,k)^2\,(1+j)+16\,M\,\omega_0\,\beta\,(2+j)+14\,M^2\,\omega^{2}_0\,\beta^2-44\,j-32\,M\,\gamma-8.
\label{aa22}
\end{eqnarray}
And that the corresponding ground state wave function given by
\begin{equation}
\psi_{1,\ell} (x)=x^{\frac{1}{4}+\frac{\sqrt{2\,M\,\gamma+\frac{1}{4}}}{2}}\,e^{-\frac{1}{2}\,M\,\omega_0\,\beta\,x}\,(c_0+c_1\,x),
\label{aa23}
\end{equation}
where $j=\sqrt{2M\gamma+\frac{1}{4}}$ and
\begin{eqnarray}
c_1=\frac{1}{4\,(1+j)}\,\Bigg[(\ell-\Phi-\beta\,k)^2-2\,M\,\omega_0\,\beta\,(j+3)-j-\frac{5}{2}\mp \sqrt{\triangle} \Bigg]\,c_0.
\label{aa24}
\end{eqnarray}

The energy eigenvalue solution of the harmonic oscillator displays intricate dependencies arising from multiple factors, namely the presence of a screw dislocation parameterized by $\beta$, the constant angular velocity of a rotating frame, $\Omega$, and the influence of a quantum flux field. These factors contribute to a shift in the angular quantum number, denoted as $\ell$, which is transformed to an effective quantum number $\ell \to \ell'=\Big(\ell-\frac{e\,\Phi_{AB}}{2\,\pi}-\beta\,k\Big)$. Notably, an inverse square potential further contributes to the modified eigenvalue solution.

One intriguing observation is that the energy eigenvalues satisfy a distinctive relationship: $E_{1,\ell} (\Phi_{AB} \pm \Phi_0\,\nu)=E_{1,\ell\mp \nu} (\Phi_{AB})$, where $\nu=0,1,2,3...$. This relationship manifests as a periodic function of the geometric quantum phase, with a fundamental periodicity represented by $\Phi_0$. This periodic dependence on the geometric quantum phase draws a striking parallel to the renowned Aharonov-Bohm effect \cite{YA,MP}, which arises in electromagnetic contexts and embodies similar periodic characteristics. Extending this framework, it is feasible to derive energy levels beyond the first state, including $E_{2,\ell}, E_{3,\ell},...$, as well as radial wave functions denoted by $\psi_{1,\ell}, \psi_{2,\ell},....$ for the radial mode with $n \geq 2$, through analogous procedures.

\section{Non-relativistic particles under flux field with inverse square potential in topological defect geometry}

In this section, we study the non-relativistic particles under the AB-flux field in topological defect with screw dislocation in the presence of an inverse square potential $V(r)=\frac{\gamma}{r^2}$. We consider the effects of rotating frame of reference in the quantum system and discuss the effects.

Therefore, the wave equation (\ref{aa4}) in the space-time background (\ref{aa3}), and using electromagnetic potential (\ref{aa2}) and inverse square potential $V(r)=\frac{\gamma}{r^2}$ under the rotational frame effects become
\begin{eqnarray}
&&-\frac{1}{2\,M}\,\Bigg[\frac{\partial^2}{\partial\,r^2}+\Big(\frac{r}{r^2-\beta^2}\Big)\,\frac{\partial}{\partial\,r}+\frac{1}{(r^2-\beta^2)}\,\Big(\frac{\partial}{\partial\,\phi}-i\,\Phi-\beta\,\frac{\partial}{\partial\,z}\Big)^2 \Bigg]\,\Psi+i\,\Omega\,\Big(\frac{\partial}{\partial\,\phi}-i\,\Phi-\beta\,\frac{\partial}{\partial\,z}\Big)\,\Psi\nonumber\\
&&+\frac{\gamma}{r^2}\,\Psi=E\,\Psi,
\label{gg}
\end{eqnarray}
Finally, substituting the wave function (\ref{aa7}) in the Eq. (\ref{gg}), we obtain the radial wave equation as follows:
\begin{equation}
\psi''+\frac{r}{(r^2-\beta^2)}\,\psi'+\Bigg[\Theta-\frac{2\,M\,\gamma}{r^2}-\frac{\iota^2}{(r^2-\beta^2)}\Bigg]\,\psi=0.
\label{gg1}
\end{equation}
where $\Theta=2\,M\,(E+\Omega\,\iota)-k^2$ and $\iota$ is given in Eq. (\ref{aa9}).

Performing a change of variable $x=\frac{r^2}{\beta^2}$ in Eq. (\ref{gg1}), we have
\begin{eqnarray}
4\,x\,\psi''(x)+\frac{4\,x-2}{(x-1)}\,\psi'(x)+\Bigg[\Theta\,\beta^2-\frac{2\,M\,\gamma}{x}-\frac{\iota^2}{(x-1)} \Bigg]\,\psi(x)=0.
\label{gg2}
\end{eqnarray}
Suppose, a possible solution to the above Eq. (\ref{gg2}) given by
\begin{eqnarray}
\psi(x)=x^{\frac{1}{4}+\frac{j}{2}}G(x),
\label{gg3}
\end{eqnarray}
where $j$ is defined earlier

Substituting solution (\ref{gg3}) into the Eq. (\ref{gg2}), once again one will arrive at the confluent Heun differential equation form \cite{AR,AR2,AR3}. Hence, $G(x)$ is the confluent Heun equation and in our case here, it is given by
\begin{eqnarray}
G\left(x\right)=H_{c}\left(0,j,-\frac{1}{2},\frac{\Theta\,\beta^{2}}{4},\frac{3}{8}-\frac{\iota^{2}+\Theta\,\beta^{2}}{4};x\right).
\label{gg4}
\end{eqnarray}

As considered earlier, we assume the function $G (x)$ to be a power series solution (\ref{aa13}), we obtain the following recurrence relation
\begin{eqnarray}
c_{j+2}=\frac{1}{h_{3}}\,(h_1\,c_{i+1}+h_2\,c_{i})
\label{gg5}
\end{eqnarray}
with the coefficient
\begin{eqnarray}
c_{1}=\frac{\Big[j+\frac{1}{2}-\Theta\,\beta^2-\iota^2\Big]}{4\left(1+j\right)}\,c_{0},
\label{gg6}
\end{eqnarray}
where we have set the parameters
\begin{eqnarray}
h_{1}=\left(i+j+\frac{3}{2}\right)\left(i+1\right)-\frac{\Theta\beta^{2}+\iota^{2}-\frac{1}{2}-j}{4},\quad h_{2}=\frac{\Theta\beta^{2}}{4},\quad h_{3}=\left(i+2+j\right)\left(i+2\right).
\label{gg9}
\end{eqnarray}

As stated earlier, to obtain the individual energy levels and wave functions, let us consider the case $i=n-1$ where the coefficient $c_{n+1}=0$. Thereby, using this condition, we obtain
\begin{eqnarray}
c_{n}=-\frac{\Theta\beta^{2}}{\left(4n\left(n-1+j+\frac{3}{2}\right)-\Theta\beta^{2}-\iota^{2}+\frac{1}{2}+j\right)}c_{n-1}.
\label{gg10}
\end{eqnarray}

Now, we can evaluate the energy levels and wave function one by one as done earlier. As particular case, the lowest state quantum system is defined by $n=1$, we obtain the coefficient
\begin{eqnarray}
c_{1}=-\frac{\Theta\beta^{2}}{\left(5j-\Theta\beta^{2}-\iota^{2}+\frac{13}{2}\right)}\,c_{0}.
\label{gg11}
\end{eqnarray}

Comparing Eq. (\ref{gg11}) with Eq. (\ref{gg6}), we obtain the ground state energy levels $E_{1,\ell}$ as follows
\begin{eqnarray}
E_{1,\ell}=\frac{k^{2}}{2\,M}+\frac{\Theta}{2\,M}-\Omega\,(\ell-\Phi-\beta\,k),
\label{gg12}
\end{eqnarray}
where we have defined
\begin{eqnarray}
\Theta=\frac{1}{\beta^{2}}\,\Bigg[j+\frac{3}{2}-(\ell-\Phi-\beta\,k)^2\pm\sqrt{(\ell-\Phi-\beta\,k)^2\,\Big(j+\frac{1}{4}\Big)-j\,\Big(j+\frac{3}{2}\Big)-\frac{1}{4}}\Bigg].
\label{gg13}
\end{eqnarray}
And that the corresponding ground state wave function is given by
\begin{eqnarray}
\psi_{1,\ell} (x)=x^{\frac{1}{4}+\frac{\sqrt{2M\gamma+\frac{1}{4}}}{2}}\,(c_0+c_1\,x),
\label{gg14}
\end{eqnarray}
where $j=\sqrt{2M\gamma+\frac{1}{4}}$ and
\begin{eqnarray}
c_1=\frac{1}{4\,(1+j)}\,\Bigg[-1 \mp \sqrt{(\ell-\Phi-\beta\,k)^2\,\Big(j+\frac{1}{4}\Big)-j\,\Big(j+\frac{3}{2}\Big)-\frac{1}{4}}\Bigg]\,c_0.
\label{gg15}
\end{eqnarray}

Equations (\ref{gg12})--(\ref{gg15}) capture the ground state energy level and the radial wave function that characterize the behavior of non-relativistic particles within a space-time environment featuring a screw dislocation. This behavior unfolds under the combined influence of a rotating frame of reference, an inverse square potential, and the AB-flux field. An insightful observation is that the presence of the topological defect, represented by the parameter $\beta$, the constant angular velocity of the rotating frame, denoted by $\Omega$, and the quantum flux field $\Phi_{AB}$ collectively exert a discernible impact on the eigenvalue solution for the non-relativistic particles. Notably, this dependence of the eigenvalue solution on the geometric quantum phase gives us the electromagnetic phenomenon known as the Aharonov-Bohm effect, as discussed in the references \cite{YA,MP}. Continuing along a similar pathway, one can determine the energy levels, such as $E_{2,\ell}, E_{3,\ell},...$, and the corresponding radial wave functions $\psi_{2,\ell}, \psi_{3,\ell},....$ for the mode $n \geq 2$.

\section{Conclusions}

To sum up, we studied the rotational frame effects on the non-relativistic particles under the influence of the quantum flux field with potential in a topological defect background with screw dislocation. We derived the radial Schr\"{o}dinger wave equation and obtained the eigenvalue solutions analytically. We have shown that the eigenvalue solution are influenced by the screw dislocation parameter characterized by $\beta$, the quantum flux field $\Phi_{AB}$, and the uniform angular speed $\Omega$ of the rotating frame with chosen potential and get them modified. One can see that the angular quantum number $\ell$ is shifted, that is $\ell \to \ell'=(\ell-\frac{e\,\Phi_{AB}}{2\,\pi}-\beta\,k)$, an effective angular momentum quantum number. This effective quantum number depends on the screw dislocation parameter $\beta$ and the quantum flux field $\Phi_{AB}$ that shows an electromagnetic analogue of the Aharonov-Bohm effect \cite{YA,MP}.

In {\tt section 2}, we have chosen a harmonic oscillator plus inverse square potential and derived the radial equation under the rotational frame effects. We presented the ground state energy level $E_{1,\ell}$ and radial wave function $\psi_{1,\ell}$ defined by the radial mode $n=1$ as particular cases. In fact, it has shown that the eigenvalue solution gets modified by the inverse square potential and the quantum flux field compared to the result obtained for the harmonic oscillator in Ref. \cite{LCNS}. In {\tt section 3}, we have chosen inverse square potential and the quantum flux field into the non-relativistic quantum system and following the same procedure presented the ground state energy levels $E_{1,\ell}$ and radial wave function $\psi_{1,\ell}$ as particular cases for the mode $n=1$. It is worth mentioning that the geometrical approach used to describe space-time with a screw dislocation can be useful in studies of condensed matter systems.

\section*{Acknowledgement}

P.R. acknowledges the Inter-University Centre for Astronomy and Astrophysics (IUCAA), Pune, India for granting visiting associateship. The authors thank the anonymous referee for his/her invaluable comments that helped them to improve the quality of the manuscript.

\section*{Data Availability}

No new data were generated in this paper.

\section*{Conflict of Interest}

There are no conflicts of interest.



\end{document}